# Integrated lithium niobate photonic millimeter-wave radar

Sha Zhu[1,†], Yiwen Zhang[2,†], Jiaxue Feng[3], Yongji Wang[4], Kunpeng Zhai[5], Hanke Feng[2,], Edwin Yue Bun Pun[2,*], Ning Hua Zhu[1,*], and Cheng Wang[2,*]

[1]*Institute of Intelligent Photonics, Nankai University, Tianjin 300071, China*
[2]*Department of Electrical Engineering & State Key Laboratory of Terahertz and Millimeter Waves, City University of Hong Kong, Kowloon, Hong Kong, China*
[3]*College of Microelectronics, Faculty of Information Technology, Beijing University of Technology, Beijing, 100124, China*
[4]*Department of Chemistry, City University of Hong Kong, Kowloon, Hong Kong, China*
[5]*State Key Laboratory on Integrated Optoelectronics, Institute of Semiconductors, Chinese Academy of Sciences, Beijing, 100083, China*
[†]*These authors contributed equally to this work.*
*Corresponding authors: eeeybpun@cityu.edu.hk, nhzhu@semi.ac.cn, cwang257@cityu.edu.hk*

**Abstract:** Millimeter-wave (mmWave, > 30 GHz) radars are key enabler in the upcoming 6G era for high-resolution sensing of the surroundings and detection of targets. Photonic radar provides an effective approach to overcome the limitations of electronic radars thanks to the high frequency, broad bandwidth, and excellent reconfigurability of photonic systems. However, conventional photonic radars are mostly realized in tabletop systems composed of bulky discrete components, whereas the more compact integrated photonic radars are difficult to reach the mmWave bands due to the unsatisfactory bandwidths and signal integrity of the underlying electro-optic modulators. Here, we overcome these challenges and demonstrate a centimeter-resolution integrated photonic radar operating in the mmWave V band (40-50 GHz) based on a 4-inch wafer-scale thin-film lithium niobate (TFLN) technology. The fabricated TFLN mmWave photonic integrated circuit consists of a first electro-optic modulator capable of generating a broadband linear frequency modulated mmWave radar waveform through optical frequency multiplication of a low-frequency input signal, and a second electro-optic modulator responsible for frequency de-chirp of the received reflected echo wave, therefore greatly relieving the bandwidth requirements for the digital-to-analog converter in the transmitter and analog-to-digital converter in the receiver. Thanks to the absence of optical and electrical filters in the system, our integrated photonic mmWave radar features continuous on-demand tunability of the center frequency and bandwidth, currently only limited by the bandwidths of electrical amplifiers. By operating on the mmWave V band 40-50 GHz, we achieve multi-target ranging with a resolution of 1.50 cm and velocity measurement with a resolution of 0.067 m/s. Furthermore, we construct an inverse synthetic aperture radar (ISAR) and successfully demonstrate the imaging of targets with various shapes and postures with a two-dimensional resolution of 1.50 cm × 1.06 cm. Our integrated TFLN photonic mmWave radar chip provides a compact and cost-effective solution in the 6G era for high-resolution sensing and detection in vehicle radar, airborne radar, and smart homes.

## Main

For decades, radio detection and ranging (radar) at microwave frequencies has been the fundamental technology for various applications, such as airborne object detection, weather forecasting, resource exploration, and vital-sign monitoring[1]. In the forthcoming 6G era, millimeter-wave (mmWave) radars operating at even higher frequencies above 30 GHz and with broader bandwidths are anticipated to play a pivotal role in Integrated Sensing and Communication (ISAC) systems that require high-resolution detection and real-time situational awareness, enabling new application scenarios like indoor sensing, automated driving, and vital-sign monitoring[2-4]. However, the operation frequency and bandwidth of traditional electrical radar systems are typically limited and often traded off with each other, making it challenging to precisely locate, recognize, and image objects with large detection ranges and fine resolution simultaneously.

The emergence of photonics-based radar technology provides a promising solution to overcome these limitations by processing microwave signals in the optical domain. This leverages the benefits of photonics systems, including high frequency, large bandwidth, low transmission loss, reconfigurability, and anti-electromagnetic interference[5-7]. In 2014, the first photonic-assisted fully digital radar (PHODIR) was demonstrated based on a mode-locked laser and optical domain frequency conversion technology[8]. A phase-coded microwave radar waveform is generated for the detection of a non-cooperative aircraft with a ranging resolution of 23 m, which was significantly limited by the narrow signal bandwidth of 200 MHz. To further improve the resolution, different architectures of microwave photonic radars have been proposed, such as frequency multiplication[9-15], optical injection of semiconductor lasers[16], photonic stretch processing[17], cyclic frequency shift[18], and frequency up-conversion[19-21] (see Extended Data Table 1). However, most microwave photonic radars to date are still constructed using discrete optoelectronic devices with significant disadvantages in terms of size, weight, power consumption, stability, and cost (SWaPSC).

Recently, integrated photonics has opened up new opportunities for improving the SWaPSC performance of microwave photonic systems by miniaturizing and integrating multiple photonic devices in chip-scale systems[22-26]. Benefiting from the compatibility with mature complementary metal-oxide semiconductor (CMOS) fabrication technology, several silicon (Si) photonic chip-based radars have been realized in the microwave S (2-4 GHz), C (4-8 GHz), X (8-12 GHz), and Ku (12-18 GHz) bands[27-29], with a maximum demonstrated bandwidth of 6 GHz (from 12 to 18 GHz) (see Extended Data Table 1). However, the modulation mechanism employed by silicon electro-optic modulators, i.e. free carrier depletion, presents inherent limitations to the achievable performances of Si photonic radars. These include restricted bandwidths, nonlinear electro-optic responses, and limited extinction ratios significantly impacting the operational frequency range of the radars and the signal quality (e.g. spurious harmonics) of their waveforms. Additionally, active silicon photonic devices are prone to two-photon and free-carrier absorption, particularly at excessive optical power levels that are often needed in microwave photonic systems[30]. As a result, although integrated Si photonic radars have achieved a relatively high level of integration, including modulators and photodetectors[29], they have not been able to reach the mmWave bands (> 30 GHz), which are highly desired in future indoor sensing, autonomous driving, as well as 6G-based imaging and sensing networks.

Thin film lithium niobate (TFLN) platform is an excellent candidate to address these challenges and bring the operation frequency of integrated photonic radars into the mmWave bands. On the one hand, TFLN exhibits a fast and linear Pockels effect, making it well suited for achieving high-speed and linear electro-optic modulators (EOMs). On the other hand, the high index contrast of TFLN allows for tight confinement of optical modes, thereby enabling the implementation of multiple photonic functionalities in a single TFLN photonic integrated circuit[31-36]. In recent years, a number of TFLN-based EOMs have been developed, achieving unprecedented performance metrics including modulation bandwidths deep into the mmWave bands[35,37-40], CMOS-compatible drive voltages[31], small footprint[41], and ultra-high modulation linearity[42]. A variety of functional photonic elements, including low-loss waveguides[43], high-quality-factor microresonators[44], waveguide crossings[45], delay lines[46], arrayed waveguide grating[47,48], acousto-optic modulators[49] etc., have also been developed. Efforts to further integrate these components into chip-scale systems through wafer-scale fabrication have led to integrated microwave photonic system (MWP) signal processors with unparalleled speed and power-consumption performances[50]. These collective achievements have paved the path for the TFLN platform to be applied to photonic mmWave radar applications that require high frequency, large bandwidth, and compact form factor at the same time.

Here, we demonstrate an integrated photonic radar system operating in the mmWave V band based on a TFLN photonic circuit, achieving centimeter-level resolutions in both ranging and velocity measurements, as well as in two-dimensional imaging. Fabricated from a 4-inch wafer-scale process, the TFLN photonic radar chip consists of a frequency multiplying module for mmWave radar waveform generation and a frequency de-chirp module for echo signal reception. Benefiting from the broad bandwidths of all photonic components in our chip and the no-filter design, the center frequency and bandwidth of the generated radar waveforms can be arbitrarily configured over a wide range, in this case 40-50 GHz limited only by the electrical amplifiers. The high carrier frequency and large bandwidth enable us to achieve multi-target ranging with a distance resolution of 1.50 cm, velocity measurement with a resolution of 0.067 m/s, and inverse synthetic aperture radar (ISAR) imaging with a two-dimensional resolution of 1.50 cm × 1.06 cm.

**Results**

Figure 1 shows the conceptual illustration and working principle of our integrated photonic mmWave radar chip. We use linear frequency modulated waveform (LFMW), whose frequency varies linearly with time, as radar waveform in this work, since it features high ranging resolution, constant modulus, Doppler tolerance, as well as a straightforward frequency de-chirp process for the echo waveforms that can greatly alleviate the sampling rate requirements of the radar receiver. To generate the mmWave LFMW signal at the transmitter side, we first implement a frequency multiplying module that performs the up-conversion of a low-frequency microwave LFMW signal. Specifically, an optical carrier with a frequency of $f_c$ is modulated by a microwave LFMW signal using a first high-speed TFLN amplitude modulator (EOM1). This fundamental LFMW signal can be relatively easily produced by a CMOS digital-to-analog converter (DAC). It features an instantaneous frequency of $f_1+kt$ that linearly changes from $f_1$ to $f_1+kT$ with a bandwidth of $B=kT$, where $T$ is the waveform period. Biasing the EOM1 at the null transmission point leads to a carrier-suppressed double sideband (CS-DSB) modulation process that projects the input microwave signal into two LFMW optical sidebands with frequencies of $f_c+f_1+kt$ and $f_c-f_1-kt$, respectively. Afterwards, the modulated optical signal is divided into two paths by a 50%:50% multimode interferometer. Optical signal in the upper path is detected by a high-speed PD (PD1) to generate a mmWave radar waveform whose initial frequency ($2f_1$) and bandwidth ($B_2=2kT=2B$) are both doubled from the DAC-input electrical signal. The generated radar waveform is then amplified by a trans-impedance amplifier (TIA) and emitted into free space by an antenna. When the emitted radar waveform encounters targets, the waveform will be reflected with a time delay of $\tau$. The reflected echo waveform is collected by a receiving antenna, amplified by a low-noise amplifier (LNA), and sent to the frequency de-chirp module consisting of a second modulator (EOM2) fabricated on the same TFLN chip. The input port of EOM2 is connected to the lower output path of the abovementioned 50%:50% multi-mode interferometer, therefore featuring two carrier frequencies of $f_c+f_1+kt$ and $f_c-f_1-kt$, which are subsequently modulated by the amplified echo signals with an instantaneous frequency of $2f_1+2kt-2k\tau$. By setting the EOM2 at the quadrature transmission point, four new optical sidebands are generated, out of which two sidebands are located in the vicinity of the two carriers, at frequencies of $f_c+f_1+kt-2k\tau$ and $f_c-f_1-kt+2k\tau$. This allows us to achieve frequency de-chirp and obtain the low-frequency target information ($2k\tau$) by beating these two relevant sidebands with the two nearby carriers at a low-speed PD (PD2) and further processing using a low-speed ADC. Finally, the ranging, velocity, and imaging information of the targets under detection are obtained through subsequent data processing (see Methods). The right panel of Fig. 1a shows an envisioned application scenario of our integrated photonic mmWave radar in future autonomous vehicles, offering high-resolution distance/velocity detection and imaging capabilities simultaneously, which are key to the enhanced safety, perception, and decision-making processes in autonomous driving.

Figure 1b shows a picture of our fabricated 4-inch LNOI wafer containing 1.50 cm × 1.50 cm dies of various passive and active components, totaling 21 pieces. The wafer was patterned by an ultraviolet (UV) stepper lithography system and dry etched by an inductively coupled plasma–reactive ion etching (ICP-RIE) system, followed by standard metallization processes (see Methods). Figure 1c shows an example TFLN photonic radar chip cleaved from a 4-inch LNOI wafer, featuring a much smaller footprint than a Hong Kong ten-dollar coin.

The fabricated TFLN EOMs exhibit measured half-wave voltages of ~ 2.8 V and 3-dB electro-optic bandwidths over 50 GHz, as shown in Fig. 1d-e, both of which are important metrics for achieving broadband and high-fidelity mmWave radar signal generation and echo processing. We first show that our photonic mmWave radar chip is capable of generating high-quality LFMW signals with arbitrarily configurable center frequency and bandwidth in the mmWave V band. Figure 1f displays the measured CS-DSB optical spectrum at the output port of EOM1 when biased at null, showing a sideband-to-carrier suppression ratio higher than 25 dB thanks to the excellent extinction ratio of TFLN EOMs. The high sideband-to-carrier suppression ratio is highly desired in practical applications to ensure a low residual fundamental component in the radar waveform, which also exceeds those reported in most microwave photonic radar systems[9,19]. After beating the CS-DSB optical signals at PD1, LFMW signals that are frequency doubled from the driving electrical signal are generated, as shown in the measured electrical spectra in Fig. 1g-h. Since no optical or electrical filters are involved in our photonic radar, the bandwidth and center frequency of the generated radar waveform could be arbitrarily chosen and continuously tuned over a broad range. Figure 1g illustrates the generated radar signals with a fixed center frequency of 45 GHz and increasing bandwidths from 2 GHz to 10 GHz with a step of 2 GHz, whereas Fig. 1h corresponds to signals with different center frequencies (41 GHz to 45 GHz, with a step of 1 GHz) and a fixed bandwidth of 10 GHz. In the rest of this paper, we choose a full 40-50 GHz range as the LFMW radar waveform to achieve the best detection resolution, which is inversely proportional to the bandwidth and center frequency of the radar waveform (see Methods). Figure 1i displays the time-domain waveform of the radar signal directly recorded using a real-time oscilloscope. The corresponding frequency-time diagram extracted from the time-domain radar waveform using short-time Fourier transform is shown in Fig. 1j. The frequency of the radar waveform linearly increases from 40 GHz to 50 GHz within a time period of 4 μs, corresponding to a linear frequency chirp of 2.5 GHz/μs, which matches well with the measured electrical spectrum.

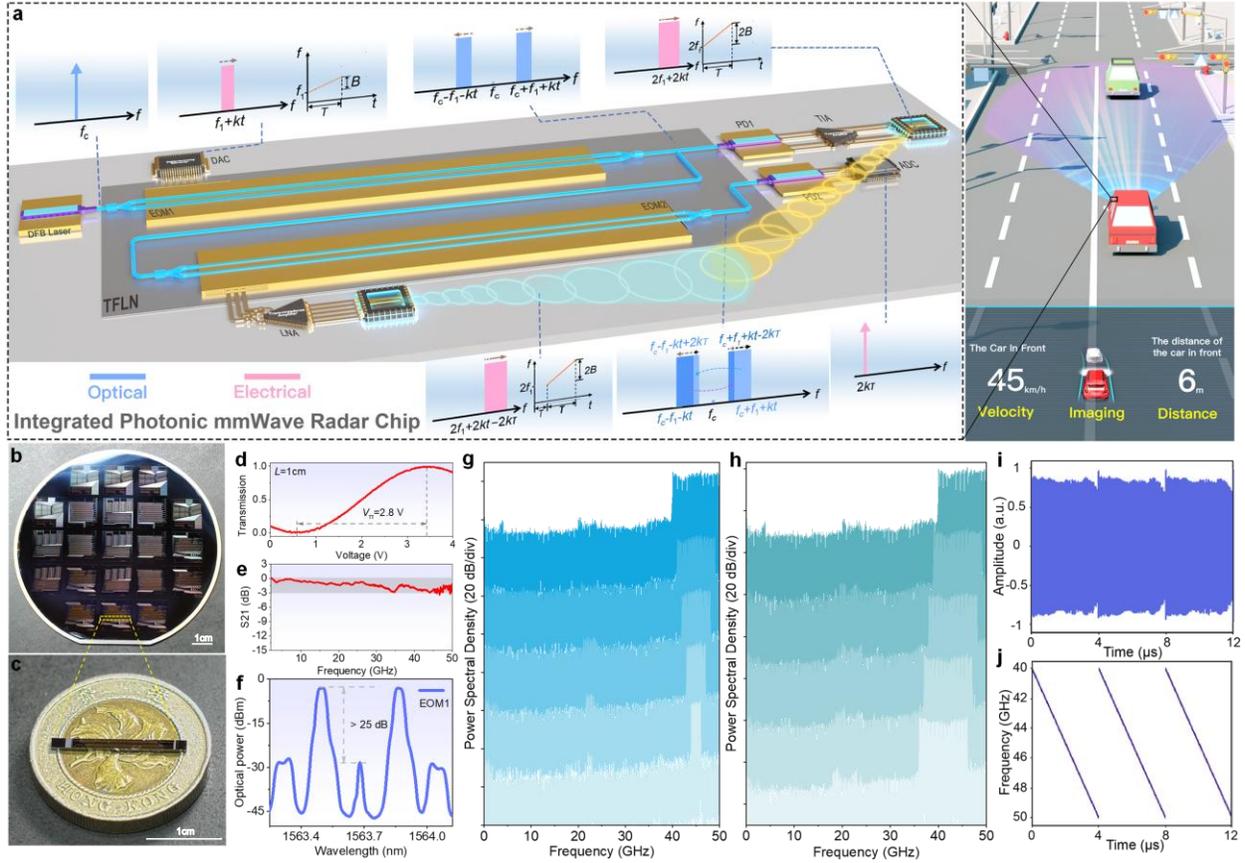

**Fig. 1. Integrated photonic mmWave radar chip and radar waveform generation. a** Conceptual drawing of an integrated photonic mmWave radar chip consisting of a first EOM that generates mmWave radar waveform via optical frequency multiplication, and a second EOM that processes the received echo signals through frequency de-chirp. Insets schematically show the frequency-domain signal spectra (blue: optical; pink: electrical) and frequency-time diagrams at different locations of the radar chip. The illustration on the right shows an envisioned application scenario where the integrated photonic mmWave radar is used for ranging, velocity detection, and imaging tasks in autonomous vehicles. **b** Picture of the fabricated 4-inch TFLN wafer. **c** Picture of the photonic mmWave radar chip cleaved from the TFLN wafer on top of a Hong Kong ten-dollar coin. **d-e** Measured electro-optic transfer function (d) and frequency response (e) of the TFLN EOM. **f** Measured optical spectra at the output of EOM1 showing > 25 dB sideband to carrier suppression ratio. **g-h** Measured spectra of the generated radar waveforms showing arbitrarily configurable bandwidths (2-10 GHz, centered at 45 GHz, g) and center frequencies (41-45 GHz, with a fixed bandwidth of 10 GHz, h). **i-j** Measured time-domain waveform (i) and frequency-time diagram (j) of the radar waveform. DFB laser: distributed feedback laser, EOM: electro-optic modulator, PD: photodetector, DAC: digital-to-analog converter, TIA: trans-impedance amplifier, LNA: low-noise amplifier, ADC: analog-to-digital converter.

**Photonic mmWave ranging radar**

We next demonstrate high-resolution ranging using our photonic mmWave radar chip. Figure 2a shows the experimental setup, the middle inset of which shows an optical microscope image of the fabricated chip. Figure 2b-c illustrates the measured ranging results for single and multiple (up to three) targets placed at various distances from the antennas, respectively. Inset $i$ of Fig. 2b shows a representative measured optical spectrum at the output of EOM2, where the two spectral peaks separated by ~ 45 GHz each include an optical carrier ($f_c$-$f_1$-$kt$ and $f_c$+$f_1$+$kt$, respectively) and a sideband generated by echo-wave modulation ($f_c$-$f_1$-$kt$+$2k\tau$ and $f_c$+$f_1$+$kt$-$2k\tau$, respectively). The closely located carriers and sidebands cannot be distinguished in the optical spectrum due to the limited resolution of the optical spectrum analyzer (OSA), but could be easily de-chirped into low-frequency and narrow-bandwidth signals at MHz level and detected using a low-frequency PD (PD2). This greatly reduces the sampling rate requirement of the ADC in the oscilloscope, which records the final time-domain waveform of the frequency de-chirped intermediate frequency (IF) signals. After a fast Fourier transform (FFT) process in real time, we obtain the electrical spectra of the de-chirped IF signals, which directly translate into the target ranges ($R_1$) (insets *ii-viii* in Fig. 2b and insets *i-v* in Fig. 2c) (see Methods). As shown in Fig. 2b and 2c, the average range resolution (3 dB bandwidth) of the ranging measurement is 1.71 cm (285 kHz), which matches well the theoretical range (frequency) resolution of 1.50 cm (250 kHz). Besides, the unwanted side lobe (the side peak closest to the highest peak) suppression ratios of the range spectra are all above 6.5 dB, indicating low microwave signal crosstalk in our chip and good detection capability for small and weak targets, which can potentially be improved by reducing system noise, increasing the number of channels, and implementing windowing and other signal processing methods[51]. In the main panels of Fig. 2b-c, we summarize and compare the measured range values and the real target distances for single-target (b) and multiple-target (c) measurements, showing accurate and linear ranging performances. The measured distance errors in single-target cases (inset *ix* of Fig. 2b) are all within ±0.15 cm for a large dynamic range of 30-420 cm, which indicates that the maximum discrepancies between the measured distances and the actual distances are all less than 0.15 cm. Our photonic mmWave radar also provides excellent ranging performances when detecting multiple targets (inset *vi* of Fig. 2c). Both two and three targets can be clearly distinguished, and the measurement distance errors are all within ±0.15 cm (inset *vii* of Fig. 2c). Importantly, we show that the photonic mmWave radar is capable of distinguishing two targets that are only 1.90 cm apart (yellow squares and inset *i* in Fig. 2c). The deviation between the measured and the theoretical ranging resolution are likely caused by echo noises from surrounding objects and air-induced transmission loss in the testing environment, resulting in a reduction of the signal-to-noise ratio (SNR) of the echo. By measuring SNR as a function of target distance and extrapolating the fitted line to zero SNR, we estimate a maximum ranging distance of 17.1 m in our current setup (See Methods and Extend Data Fig. 3).

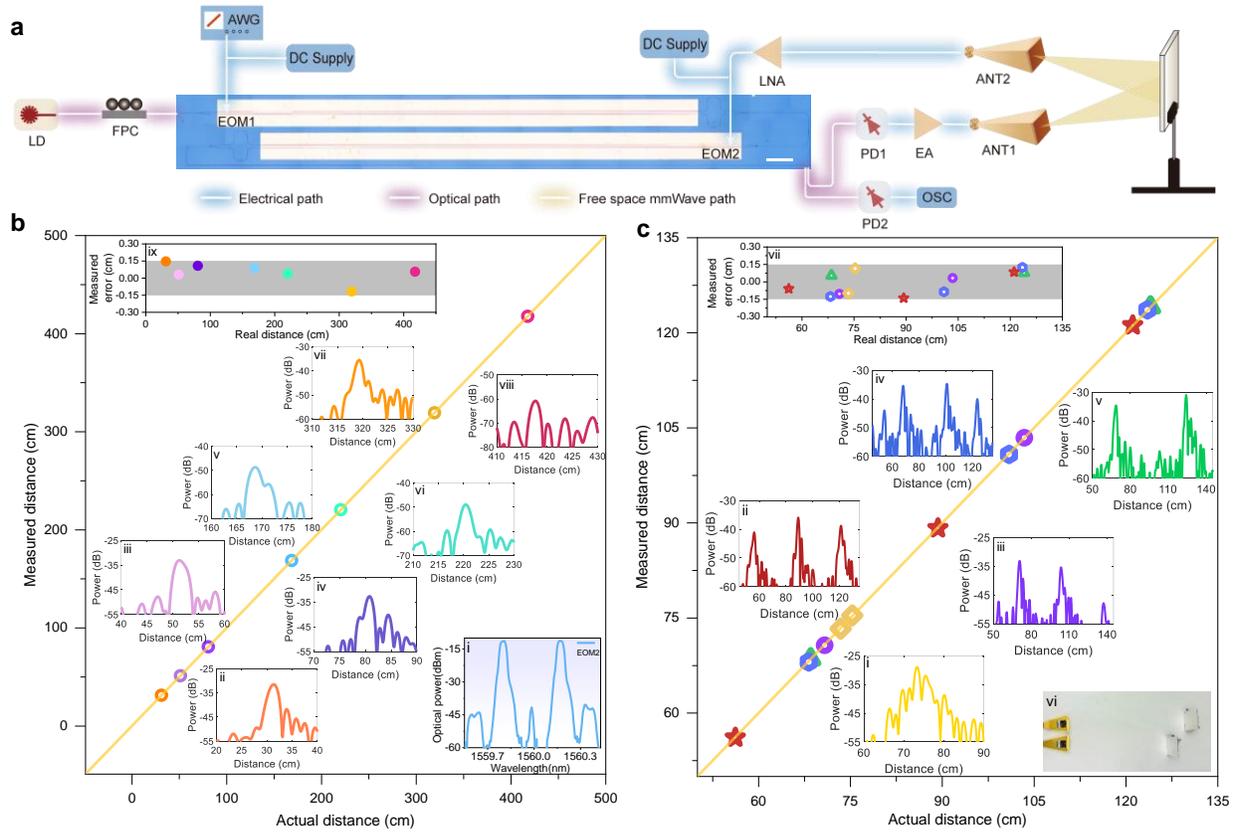

**Fig. 2. High-resolution photonic mmWave ranging radar. a.** Experimental setup for the radar ranging measurements. Inset shows a microscope image of the photonic mmWave radar chip under test. **b.** Measured distances versus actual distances for single targets placed at various locations. Insets: *i)* measured optical spectrum at the output of EOM2, *ii-viii)* measured de-chirped electrical spectra for different target distances (horizontal axes have been converted into distance values for better visualization), *ix)* measured ranging errors (the differences between measured distances and actual distances) for different target distances. **c.** Measured distances versus actual distances for two and three targets placed at various locations. Insets: *i-v)* measured de-chirped electrical spectra for the various testing scenarios, *vi)* top-down picture of the testing setup illustrating the relative positions of the antennas and targets, *vii)* measured ranging errors (the differences between measured distances and actual distances) for the various testing scenarios. Yellow lines in (b-c) correspond to the ideal relationship ($y = x$) between measured and real distances. AWG: arbitrary waveform generator, LD: laser diode, FPC: fiber polarization controller, EA: electrical amplifier, ANT: antenna, OSC: oscilloscope.

**Photonic mmWave velocity-detection radar**

Besides distance detection, our photonic mmWave radar is also capable of high-resolution velocity detection, which relies on measuring the Doppler shift in the echo signal introduced by the motion of the targets (see Methods). To demonstrate this capability, a small balanced car with tunable velocity is employed as the detection target, as depicted in the inset of Fig. 3a. The dots in Fig. 3a represent the measured velocities based on versus different set velocity values, whereas Fig. 3b shows the corresponding velocity measurement errors (which means the maximum discrepancies between the measured velocities and the set velocities are within ±0.017 m/s), revealing high velocity-detection fidelity over a wide velocity range of 0-1.2 m/s. Most notably, our photonic mmWave radar chip could successfully detect small velocities (corresponding to small Doppler shifts) down to 0.056 m/s. Further applying a two-dimensional Fourier transform to the de-chirped electrical waveforms enables simultaneous extraction of the velocity and range information of the targets, as demonstrated in Fig. 3c-j, where the vertical (velocity) span of the signals agrees well with the theoretical velocity resolution of 0.067 m/s. Currently the tested velocity range is constrained by the experimental environment and the speed limitation of the balanced car (1.5 m/s). We estimate a maximum unambiguous velocity of 833 m/s based on our current device metrics and experimental setup, which is bounded by position-velocity coupling effect due to range-Doppler coupling at high velocity (see Methods). This effect can be mitigated by implementing Doppler compensation algorithms or adopting triangular wave or dual chirp radar waveforms[52-54]. Such multi-dimensional environmental sensing capability could find applications in various domains, including automotive systems, traffic management, and object tracking.

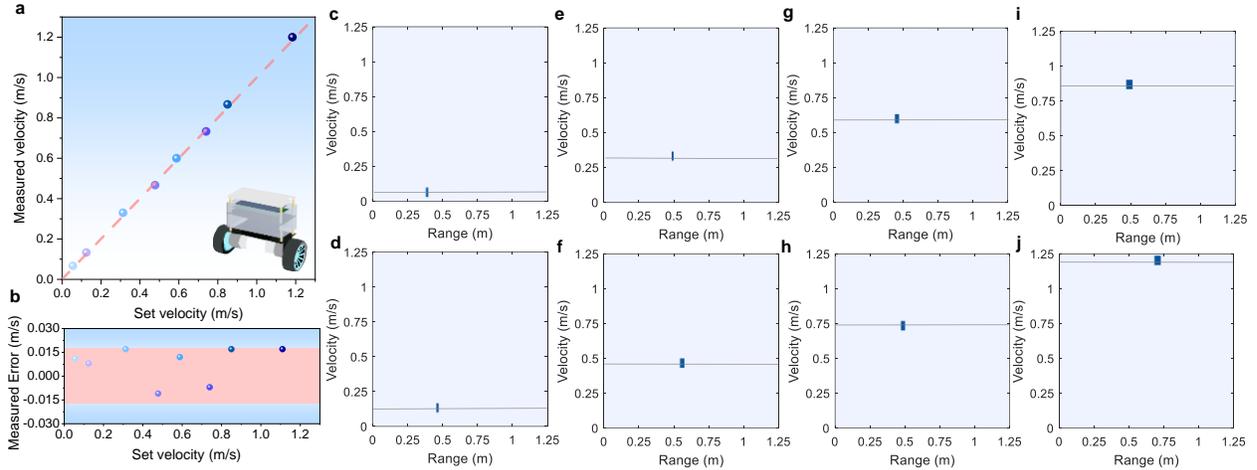

**Fig. 3. High-resolution photonic mmWave velocity-detection radar. a.** Measured target velocities at different set velocities showing marginal deviation from the ideal relationship (dashed line). Inset shows an illustration of the balanced car used as a velocity detection target. **b.** Measured velocity errors (the differences between measured velocities and set velocities) at different set velocities. **c-j** Measured two-dimensional velocity-range diagrams at various distances and velocities, the gray lines show the set velocity values.

**Photonic mmWave inverse synthetic aperture radar (ISAR) imaging**

Finally, we show the photonic mmWave radar could support high-resolution imaging tasks by constructing an ISAR as shown in Fig. 4a, where the transceiver antennas are fixed while the target undergoes a rotation of 1 round per second in the horizontal plane during detection (see Methods for more details). To characterize the basic imaging ability of the photonic mmWave ISAR, we first place several small metallic corner reflectors with a size of 3 cm × 3 cm (inset *i* of Fig. 4b) in different arrangements (Fig. 4b) on a turntable. The resulting images (insets *ii-iv* of Fig. 4b) clearly reveal the numbers, sizes, and relative locations of the corresponding metal plates in good agreement with the actual settings as indicated in the upper left insets in each case. The radar images here and in subsequent panels of Fig. 4 represent top-down views of the targets under imaging, where the vertical axes represent the radial distances (ranges) from the radar antenna and the horizontal axes correspond to the azimuthal locations (crossranges). The results show that our photonic mmWave ISAR is able to resolve and image multiple closely spaced targets (four in inset *iv* of Fig. 4b) simultaneously at a relatively long range of ~ 1.7 m. To further demonstrate the capability of imaging real-world targets of different shapes, sizes, and poses, we replace the metal plates with a number of more complicated objects, including a large airplane (45 cm × 49 cm, Fig. 4c), a medium airplane (33 cm × 34 cm, Fig. 4d), a small airplane (21 cm × 24 cm, Fig. 4e), and a doll (30 cm × 20 cm, Fig. 4f). The processed images in Fig. 4c-f show that our system can successfully resolve the structural outlines of these targets at various rotation angles. It should be noted that the imaging signals are weaker and sometimes absent at the bottom sides of these images, since they correspond to features away from the radar antenna and the echo signals from these areas could be blocked by other thicker parts of the target. Nevertheless, our results show the clear distinction of tiny features like the 5 cm empennage of the small airplane (inset *ii* of Fig. 4e) and the 0.7 cm wide arms and legs of the doll (Fig. 4f), proving the successful achievement of a centimeter-resolution photonic mmWave ISAR using our integrated TFLN chip. To quantitatively evaluate the azimuthal resolution of our ISAR, we perform imaging of a single metal corner reflector and measure the azimuthal signal intensity distribution, which yields a 3-dB azimuthal resolution of 1.06 cm, consistent with the theoretical value (Extended Data Fig. 4). The slight defocusing in our ISAR two-dimensional images can be attributed to the noises from a non-darkroom environment and reflectance variations in different parts of the targets, which could be improved using imaging algorithms with translational and rotational compensation[55,56]. Although the target movement is more complicated in reality, it can be restored into a turntable model after translational motion compensation.

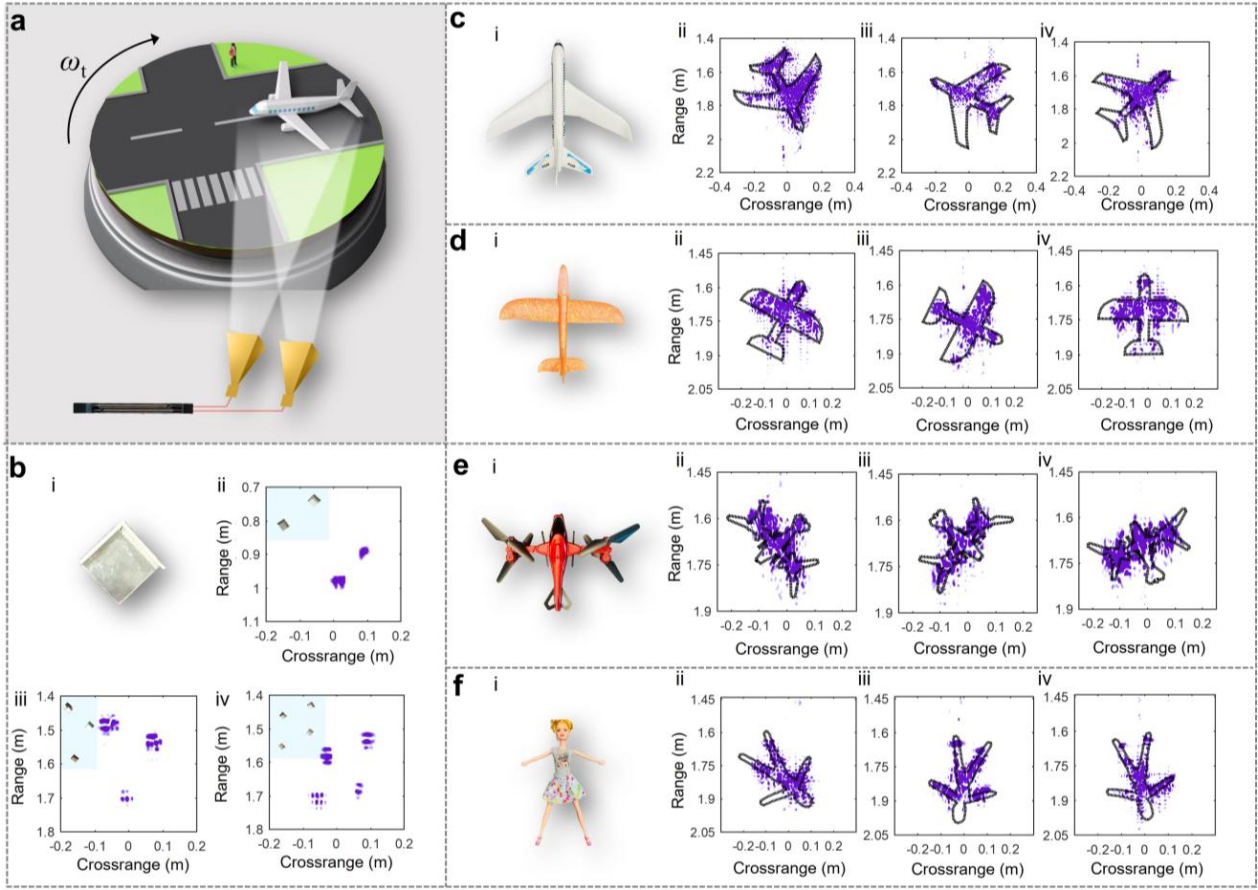

**Fig. 4. High-resolution photonic mmWave imaging radar. a.** Schematic illustration of the imaging radar test scene. **b.** Radar imaging results for various numbers and arrangements of small metallic corner reflectors with a size of 3 cm × 3 cm. **c-f** Radar imaging results for a large (**c**), a medium (**d**), and a small (**e**) airplane model, as well as a doll (**f**), imaged at various azimuthal rotation angles.

**Discussions**

In summary, we have designed, fabricated, and demonstrated an integrated photonic mmWave radar that is capable of high-resolution ranging, velocity measurement, and ISAR imaging tasks. The demonstrated centimeter-level resolutions are enabled by the high operation frequency and broad bandwidth (40-50 GHz) of our emission mmWave LFMW generated from an on-chip optical frequency multiplexing module. The subsequent frequency de-chirp module also significantly relieves the sampling rate requirements for the ADC and PD in the radar receiver. Benefiting from the excellent modulation performance and scalability of the TFLN platform, our results show significantly improved overall performance in terms of integration level and radar resolutions compared with previous photonic radar demonstrations (see Extended Data Table 1).

Even better radar resolution down to millimeter levels could be realized by further increasing the frequency and bandwidth of the radar waveforms into upper and more practically important mmWave bands (e.g. 76 – 79 GHz for vehicular radars). Leveraging capacitively loaded traveling-wave electrodes[37], TFLN EOMs with 3-dB bandwidths exceeding 80 GHz (extrapolated from our previous work[50]) could be realized using the same wafer-scale fabrication platform. Furthermore, more advanced frequency multiplying architectures could be implemented to maintain a relatively low bandwidth requirement for the DAC. For example, frequency quadrupling could be achieved through beating the second-order sidebands, by biasing EOM1 at the full transmission point (to suppress the first-order sidebands) and incorporating a micro-resonator to filter out the optical carrier signal. Replacing EOM1 with a dual-parallel Mach-Zehnder modulator could potentially generate radar waveforms with octupled frequencies and bandwidths. The excellent performance and scalability of the TFLN platform together with recent efforts in heterogeneous and hybrid integration have unlocked the possibility of integrating and co-packaging all the employed photonic and electrical devices on chip and board levels, leading to the realization of a compact and low-cost integrated photonic mmWave radar full system. This includes the integration of lasers[44,57-59], amplifiers[44,60], PDs[59,61], and CMOS electronic components[62] on the LNOI platform through rare earth ion doping or heterogeneous integration. High-frequency mmWave antennas could potentially be fabricated directly on the TFLN chip benefiting from the reduced wavelengths at mmWave frequencies[63]. The low half-wave voltage of TFLN modulators is also a critical enabler towards compact optoelectronic co-packaging between the DAC chip and the TFLN chip, without the need of an electrical amplifier in between. The TFLN photonic mmWave radar could provide compact, low-cost and high-resolution solutions for diverse applications such as sensing[64], imaging[29], smart home[65], environmental monitoring[66] and the seamless integration of communication and radar systems in the anticipated 6G technology era[22].

## Methods

### Design and fabrication of the devices:

Devices are simulated using Ansys Lumerical Mode and High Frequency Simulation Software (Ansys HFSS). A 4-inch TFLN wafer with a 500 nm x-cut device layer from NANOLN is used to fabricate the designed devices. First, an etching hard mask of $SiO_2$ is deposited on the LNOI surface through plasma-enhanced chemical vapor deposition (PECVD). An ASML UV Stepper lithography system (NFF, HKUST) with a resolution of 500 nm patterns waveguides, EOMs, and MMI on the 4-inch LNOI wafer die by die (1.5 cm×1.5 cm). Then, the exposed pattern is transferred to the $SiO_2$ hard mask using a standard fluorine-based dry etching process and to the TFLN layer with 250 nm waveguide height and 250 nm slab height using an optimized $Ar^+$ assistant ICP-RIE process. Afterwards, a second stepper lithography patterns the electrode layer after removing the residual $SiO_2$ mask. After metal evaporation and lift-off process, ground-signal-ground electrodes with a gap of 5.5 µm are obtained which can ensure good electro-optic modulation efficiency and low metal-induced optical losses. Finally, the chips are cladded in $SiO_2$ using PECVD, cleaved and facet polished carefully.

### Principles of the integrated photonic mmWave radar:

Here, the detailed principles of the integrated photonic mmWave radar are presented. In order to generate the radar waveform, the EOM1 is driven by a fundamental linear frequency modulated signal. The output optical signal at the output of EOM1 is expressed as

$$E_1(t) = \frac{1}{2}E_0 e^{j2\pi f_c t} \cdot (e^{j\frac{\pi V_1}{V_\pi}\cos(2\pi(f_1+\frac{1}{2}kt)t)} + e^{-j\frac{\pi V_1}{V_\pi}\cos(2\pi(f_1+\frac{1}{2}kt)t)} \cdot e^{j\frac{\pi V_{DC1}}{V_\pi}}) , (1)$$

where $E_0$ and $f_c$ are the amplitude and frequency of the optical carrier, $V_1$, $f_1+kt$, $k$, $T$ and $B$ ($B=kT$) are the amplitude, instantaneous frequency, chirp rate, waveform period, and bandwidth of the fundamental linear frequency modulated signal, $V_{DC1}$ is the applied DC bias voltage and $V_\pi$ is the half-wave voltage of the EOM1. By setting the DC bias of EOM1 at the null transmission point, a carrier-suppressed double-sideband (CS-DSB) modulated optical signal is obtained which can be written as

$$E_1(t) = E_0 J_1(\beta_1) \cdot (e^{j[2\pi(f_c+f_1+\frac{1}{2}kt)\cdot t+\frac{\pi}{2}]} + e^{j[2\pi(f_c-f_1-\frac{1}{2}kt)\cdot t+\frac{\pi}{2}]}) , (2)$$

where $J_n$ is the $n$-order first-kind Bessel functions, $\beta_1 = \pi V_1/V_\pi$ is the corresponding microwave signal modulation index of the EOM1. Afterwards, the optical signal is divided into two parts by a one-in-two (50%:50%) multimode interferometer. As shown in the upper path of Fig. 1a, half the optical signal is detected by PD1 to achieve photoelectric conversion. The recovered high-frequency electrical signal in PD1 can be given by

$$I_1(t) \propto 2E_0^2 J_1^2(\beta_1) \cdot \cos(2\pi(2f_1+kt)\cdot t) . (3)$$

Thus, the initial frequency ($f_2=2f_1$) and bandwidth ($B_2=2kT=2B$) of the generated radar waveform are doubled from the driving electrical signal. Afterwards, the radar waveform is amplified by a TIA and emitted to free space by an antenna. When the emitted radar waveform meets a target, the waveform will be reflected with a time delay of $\tau$. The reflected echo waveform is collected by a received antenna and amplified by a low-noise amplifier (LNA). By applying the amplified echo waveform to EOM2 and biasing EOM2 at the quadrature transmission point, the optical signal at the output of EOM2 is expressed as

$$E_2(t) = \sqrt{2}E_0 J_1(\beta_1)J_0(\beta_2) \cdot (e^{j[2\pi(f_c+f_1+\frac{1}{2}kt)\cdot t+\frac{3\pi}{4}]} + e^{j[2\pi(f_c-f_1-\frac{1}{2}kt)\cdot t+\frac{3\pi}{4}]})$$

$$+\sqrt{2}E_0 J_1(\beta_1)J_1(\beta_2) \cdot (e^{j[2\pi(f_c+3f_1+\frac{3}{2}kt-2k\tau)\cdot t+\frac{3\pi}{4}-2\pi(2f_1-k\tau)\cdot\tau]} + e^{j[2\pi(f_c-f_1-\frac{1}{2}kt+2k\tau)\cdot t+\frac{3\pi}{4}+2\pi(2f_1-k\tau)\cdot\tau]} , (4)$$

$$+e^{j[2\pi(f_c+f_1+\frac{1}{2}kt-2k\tau)\cdot t+\frac{3\pi}{4}-2\pi(2f_1-k\tau)\cdot\tau]} + e^{j[2\pi(f_c-3f_1-\frac{3}{2}kt+2k\tau)\cdot t+\frac{3\pi}{4}+2\pi(2f_1-k\tau)\cdot\tau]})$$

where $\beta_2 = \pi V_2/V_\pi$ and $V_2$ are the corresponding modulation index of the EOM2 and amplitude of the echo waveform.

Afterwards, the PD2 is used to recover the frequency de-chirped electrical signal, by beating the two optical sidebands at frequencies of $f_c+f_1+kt$ and $f_c+f_1+kt-2k\tau$ (or $f_c-f_1-kt+2k\tau$ and $f_c-f_1-kt$), which is given by

$$I_{de}(t) \propto 4J_0(\beta_2)J_1(\beta_2)\cos\left[2\pi\cdot(2k\tau)\cdot t+2\pi(2f_1-k\tau)\tau\right] . (5)$$

The frequency ($f_{de}$) of the de-chirped electrical signal is $2k\tau$, which is proportional to the time delay of the echo waveform. Thus, the range ($R_1$) between the radar antenna and the detected target can be equal to c$\tau$/2, which is written as

$$R_1 = \frac{cTf_{de}}{4B} . (6)$$

Therefore, the ranging detection can be achieved. If $f_{de}$ is equal to the minimum distinguishable frequency difference ($f_{min}=1/T$), the detected range is called theoretical ranging resolution ($\Delta R$) which can be expressed as

$$\Delta R = \frac{c}{4B} = \frac{c}{2B_2} . (7)$$

According to the radar equation, the distance between target and radar $R$ can be expressed as

$$R = (\frac{P_t G_t \sigma A_e}{(4\pi)^2 P_r})^{1/4} , (8)$$

where $P_r$ is the power received by the radar, $P_t$ is the peak transmitter power, $G_t$ is the transmitter gain, $\sigma$ is the radar cross section, $A_e$ is he effective area of the receiving antenna. The angular resolution, defined as the distance between two targets, can be calculated using the following formula as:

$$S_A \geq 2R\sin(\theta/2) \text{ , (9)}$$

where $\theta$ represents the antenna beamwidth from commercial antenna datasheet, $S_A$ is the angular resolution as a distance between two targets, and $R$ is the slant range aim.

Additionally, our radar possesses velocity detection capabilities. When a target moves towards or away from the radar line of sight by a distance of $\Delta r$, the received signal experiences an additional delay ($\Delta\tau$) which is introduced by Doppler shift compared to the echo signal received when the object is stationary. This additional phase change in the de-chirped electrical signal can be expressed as

$$\Delta\varphi = 2\pi f_{rc}\Delta\tau = \frac{4\pi\Delta r}{\lambda} = \frac{4\pi vT}{\lambda} \text{ , (10)}$$

where $f_{rc}$ is the center frequency of the radar waveform, $\lambda$ is the wavelength of de-chirped electrical signal, $T$ is the period of the chirped transmission signal. The phase exhibits a linear response to small distance changes in the object. The phase difference measured across two consecutive chirps can be used to estimate the velocity of the target. Hence, the estimated velocity, obtained from the measured phase difference, is expressed as

$$v = \frac{\lambda\Delta\varphi}{4\pi T} \text{ , (11)}$$

Based on the properties of the Fourier transform, the peak phase of the Fourier spectrum represents the initial phase of the signal. However, when measuring velocity with multiple objects using $M$-period chirps transmitted within a 'frame' ($T_f=1/MT$), the values at the peak that contain phasor components from different targets can be distinguished by performing an FFT on the sequence corresponding to the range-FFT peaks, commonly known as a Doppler-FFT. If $\Delta\varphi$ equals the minimum distinguishable phase difference ($\Delta\varphi_{min}=2\pi/M$), the detected velocity is referred to as the theoretical velocity resolution ($\Delta v$), which can be calculated as

$$\Delta v = \frac{\lambda}{2MT} = \frac{\lambda}{2T_f} = \frac{c}{2T_f f_{rc}} \text{  (12)}$$

The maximum unambiguous velocity can be calculated as

$$v_m = \frac{c \cdot f_{PRF}}{2f_2} \text{ , (13)}$$

where $f_{PRF}$ is the repetition frequency of the radar waveform.

Finally, an ISAR system has been demonstrated using our integrated photonic mmWave radar chip. A turntable model is used to analyze and simplify the movement of targets for ISAR imaging. The imaging process involves the relative rotation between the radar and the target, which can be observed in the joint Range-Doppler domains. The target that is detected undergoes rotation at an angular velocity of $\omega_i$. To process the de-chirped electrical signal, we employ a low-speed OSC that performs multi-period sampling for a duration of $T_i$, equivalent to one coherent processing interval (CPI). The collected data is then rearranged into a two-dimensional matrix (M×N) of delay time (M) and pulse number (N). By leveraging the intrinsic relationship between delay time and distance, it is possible to achieve distance compression through the application of a Fourier transform on M data points within each received echo, thereby generating a range envelope. Consequently, the resonant peak discerned within the range envelope accurately denotes the positional distance of the principal scattering point. The distance can be calculated based on the distance formula [Eq. (6)]. The position of the amplitude peak in the distance envelope represents the distance location of the main scattering points. The translational component of the target with respect to the radar is useless for ISAR imaging, so motion compensation (including distance alignment and phase compensation) on the range compressed data to eliminate phase terms is necessarily required. At each distance unit, we conduct a Fourier transform on the reflected signals of N pulses, generating an N-point Doppler frequency domain and achieving azimuthal compression. As a result, through the initial FFT process of the 2D-FFT, the complex information obtained is equivalent to that acquired through physical I/Q mixing[67]. By applying Fourier transforms to both the distance and azimuthal dimensions, we can obtain the image of the measured target. The resolution in the distance dimension remains the same as that in ranging detection, while the resolution in the azimuthal dimension ($\Delta A$) is expressed as

$$\Delta A = \frac{cf_a}{2f_{rc}\omega_t} = \frac{c}{2f_{rc}T_i\omega_t} = \frac{c}{2f_{rc}\Delta\theta} \text{ , (14)}$$

where $f_a$ is equal to the minimum distinguishable frequency difference ($f_{min}=1/T_i$), $f_{rc}$ is the center frequency of the radar waveform, $\Delta\theta$ is the accumulated angle during radar detection. The azimuthal resolution is determined by the total number of pulses (N) within the CPI.

**Principle of the TFLN-based frequency doubling module with limited Y-branch optical splitting ratio:**
Due to fabrication errors, the splitting ratio of the Y-branch in the intensity modulator deviates from the ideal value of 50%:50%. Here we set $g$ and $1-g$ as the optical splitting ratio of the Y-branch. By biasing the intensity modulator at the null transmission point and considering small-signal modulation, the output optical signal can be expressed as

$$E_{re1}(t)=E_0e^{j2\pi f_c t} \cdot (ge^{\frac{j\pi V_1}{V_\pi}\cos\left[2\pi(f_1+\frac{1}{2}kt)t\right]} + (1-g)\ e^{-j\frac{\pi V_1}{V_\pi}\cos\left[2\pi(f_1+\frac{1}{2}kt)t\right]} \cdot e^{j(\frac{\pi V_{DC1}}{V_\pi})}) \text{ , (15)}$$

$$=(2g-1)E_0J_0(\beta_1)e^{j2\pi f_c t} + E_0J_1(\beta_1)(e^{j[2\pi(f_c+f_1+\frac{1}{2}kt)t+\frac{\pi}{2}]} + e^{j[2\pi(f_c-f_1-\frac{1}{2}kt)t+\frac{\pi}{2}]})$$

where $E_0$ and $f_c$ are the amplitude and frequency of the optical carrier, $V_1$, $f_1+kt$ and $k$ are the amplitude, instantaneous frequency and chirp rate of the applied LFMW, $V_\pi$ is the half-wave voltage of the modulator, $V_{DC1}=V_\pi$ is the applied DC bias voltage, $J_n$ is the $n$-th order Bessel functions of the first kind, $\beta_1=\pi V_1/V_\pi$ is the corresponding RF modulation index. As shown in Eq. (15), the output consists of two first-order sidebands and the residual optical carrier. Subsequently, the optical signal is detected by a PD to achieve photoelectric conversion. The generated radar waveform is given by

$$I_{PD1}(t) \propto (2g-1)^2 E_0^2 J_1^2(\beta_1) + 2E_0^2 J_1^2(\beta_1) + 2E_0^2 J_1^2(\beta_1)\cos[2\pi(2f_1+kt)t], \text{ (16)}$$

where the first and second terms represent the direct current component, and the third term is the generated frequency doubled LFMW radar signal. Thus, the residual optical carrier does not generate harmonics. In addition, we have conducted additional simulations using MATLAB, as illustrated in the Extended Data Fig. 2. Even when the sideband-to-optical carrier ratio decreases from 24 dB to 9 dB, the fundamental frequency tone remains suppressed in the output electrical spectra. The higher-order harmonics (80-100 GHz) are automatically filtered out due to the 50 GHz bandwidth of the photodetector.

**Characterization of the devices on the TFLN platform:**
To test the optical performance of the fabricated TFLN chips, optical input signal from a tunable telecom laser source (Santec TSL-550) is coupled to the chip using a lensed fiber. The output optical signal is collected by another lensed fiber and sent to a low-speed PD (125 MHz New Focus 1811). The optical loss of the TFLN waveguides is estimated by measuring the optical transmission spectrum of a racetrack resonator and fitting with a Lorentzian function. The fabrication process of this chip is the same as our previous work and the waveguide propagation loss is < 0.1 dB/cm[50].

To measure the half-wave voltage ($V_\pi$) of the fabricated EOMs, a kilohertz electrical triangular waveform generated from an arbitrary-waveform generator (AWG, RIGOL DG4102) is applied to the ground-signal-ground electrodes of the EOM through a probe (GGB industries, 50 GHz). The output optical signal of the EOM is detected using the same low-speed PD and monitored using a low-speed oscilloscope (RIGOL DS6104). For electro-optic $S_{21}$ response measurements, a frequency sweeping electrical signal generated from a 53-GHz vector network analyzer (VNA, Keysight E5080B) is sent to the EOM with a 50 Ω load. A high-speed PD (XPDV2120R) is used to detect the modulated optical signal. The recovered electrical signal is sent back to the input port of the VNA. After calibrating the frequency responses of the probe, electrical cables, and PD, the $S_{21}$ frequency response of the EOM can be obtained, showing 3-dB bandwidths larger than 50 GHz in this case. Compared with discrete photonic radars, we not only reduce the size of the modulation block from $2 \times 135.0$ mm $\times 11.4$ mm to 15 mm $\times 1.5$ mm, but more importantly substantially improves the modulation half-wave voltage and bandwidth performances, as detailed in Fig. 1. The calculated RF $V_\pi$, of our modulator is much lower than commercial bulk lithium niobate modulator (Thorlab, LNA6213), as shown in Extended Data Fig. 1[38]. These metrics directly translate into the power consumption, frequency band, resolution and size of the final photonic radars.

**Ranging, velocity, and imaging measurement of the photonic mmWave radar:**
In our proof-of-concept radar detection experiments, a continuous wave optical carrier emitted from the tunable laser is first amplified by an erbium doped fiber amplifier (HaoMinOE EDFA-C-4) and coupled to the TFLN radar chip through a lensed fiber. Benefiting from the high-power handling capability of TFLN, the high input optical power can improve the SNR of the modulated output optical signal from the chip. A fiber-polarization-controller (FPC) is used to tune the input optical signal to TE mode for the largest EO modulation efficiency. The fundamental LFMW signal (with instantaneous frequency linearly increasing from 20 GHz to 25 GHz within 4 μs) is generated from a high-speed AWG (Keysight M8196A, 33 GHz), amplified by an electrical power amplifier (Fairview microwave PE15A4021), combined with a DC voltage through a bias-tee (Marki microwave BT-0050), and used to drive EOM1 via the high-speed probe. The DC voltage is used to bias EOM1 at the null transmission point for the CS-DSB modulation scheme here. Optical spectrum of the CS-DSB modulated signal is monitored using an OSA (YOKOGAWA AQ6370D). At the output side of the photonic mmWave radar chip, a lensed fiber array is used to collect the output optical signals from EOM1 and EOM2, respectively. The upper output path of EOM1 is amplified and detected by the high-speed PD1 to generate mmWave radar waveform, which can be monitored using an electrical spectrum analyzer (Agilent N9030A) and analyzed using a high-speed oscilloscope (Keysight UXR0404AP). In actual radar testing, the generated radar signal is amplified by two-stage electrical amplifiers (Centellax OA4MVM3) before emitted to free space through a horn antenna (SAGE Millimeter InC WR-22 SAZ-2410-22-S1). The reflected echo waveform is collected by another horn antenna of the same type, amplified first by a low-noise amplifier (SHF S807) and then a power amplifier (Centellax OA4MVM3), before used to drive EOM2 through another high-speed probe. The optical output of EOM2 is amplified and detected by the low-speed PD2. An oscilloscope (DSO-X 91604A) is used to capture the recovered electrical signal with a sampling rate of 1 GSa/s. The low-speed PD2 and oscilloscope in the receiver possesses natural filtering characteristics as it cannot respond to electrical signals beyond its bandwidth. Consequently, undesired harmonics are naturally eliminated. Additionally, a low-pass electrical filter can be incorporated into the radar receiver to further ensure effective harmonic filtering.

For ranging experiments, metallic plates with a size of 7 cm $\times$ 10 cm are used as targets, which are placed at different positions with respect to the antenna. After a fast FFT process in real time (with a latency of 40 ms in a personal computer), we can obtain the distance between the target and radar with a resolution of 1.50 cm. By integrating artificial intelligence technologies such as Convolutional Neural Networks (CNNs) with radar processing algorithms, it is possible to enhance both the refresh rate and detection accuracy even further. Additionally, upgrading the computer platform configuration can also lead to further improvements in processing speed. The detection distance of a radar system is directly proportional to the power of its emission, reflected signal intensity, radar antenna gain [Eq. (8)]. Based on the test results, we can derive a fitting curve related to distance and received power, as shown in Extended Data Fig. 3(a). The Extended Data Fig. 3(b) shows the signal to noise ratio of our measurement results during different distance between target and antenna. By extrapolating the fitted line to the point where the SNR approaches zero, we estimate the maximum detection distance of our radar to be 17.1 m under our current equipment setting, mainly limited by the electrical amplifiers, antennas, as well as environmental conditions. We could improve these points and achieve larger ranges of hundreds of meters by increasing the intensity of the radar transmitted and reflected signals. Moreover, the angular resolution can be calculated from the Eq. (9), which implies the angular resolutions are 71 cm and 81 cm on the E-plane and H-plane at range of 420 cm, respectively. For velocity measurements, a toy balanced car with a size of 9 cm $\times$ 6 cm is used as the target, whose velocity can be programmed by a field programmable gate array (FPGA). By sampling 50 ms of the recovered electrical signal using an oscilloscope, velocity information is obtained after Doppler-FFT, corresponding to a velocity resolution of 6.7 cm/s. For ISAR demonstration, the targets are placed on a home-made turntable with a set rotate speed of 1 round per second. The images of the detected targets are obtained with a two-dimensional resolution of 1.50 cm × 1.06 cm after a two-dimensional FFT process to the collected 50 ms recovered electrical signal. In order to show the azimuthal resolution clearly, we illustrated the measured ISAR results of a single metal corner reflector in Extended Data Fig. 4, which includes the ISAR two-dimensional imaging, the power distribution along an azimuthal slice and the corresponding zoomed-in views, respectively. From Extended Data Fig. 4d, it could be seen that the measured azimuthal resolution at 3 dB bandwidth is 1.06 cm.

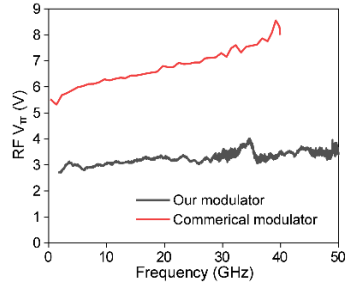

**Extended Data Fig. 1** Comparison of drive voltages as functions of frequency between a commercial bulk lithium niobate modulator (Thorlab, LNA6213) and our TFLN modulator.

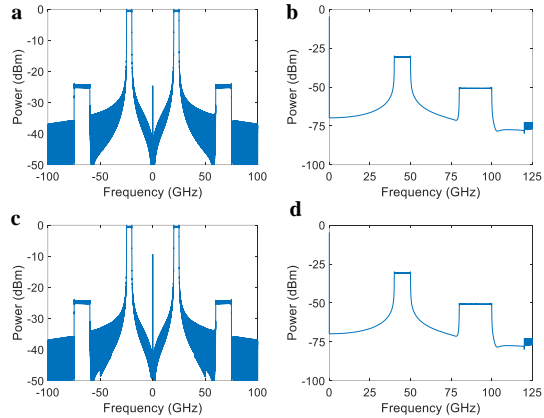

**Extended Data Fig. 2.** Simulated performance of the frequency doubling module with limited Y-branch optical splitting ratio. The **a**, **c** optical spectra and **b**, **d** electrical spectra at the output of EOM1 (a-b) and PD1 (c-d) for sideband-to-optical carrier ratio of 24 dB and 9 dB, respectively.

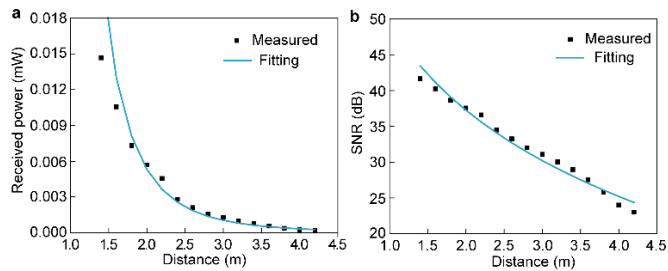

**Extended Data Fig. 3.** Echo performance of the proposed radar at different target distances. **a** Received

power of a small metal plate target (7 cm × 10 cm) versus distance between the transmitting and receiving antenna. **b** Measured (black points) signal-to-noise ratio (SNR) of received signal as a function of distance, together with fitting (blue line) results according to antenna theory.

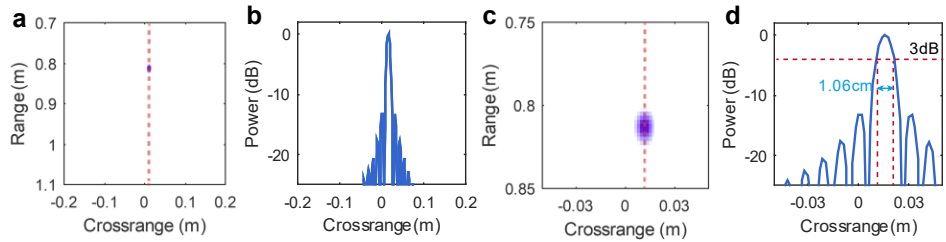

**Extended Data Fig. 4**. Azimuthal resolution of the proposed ISAR imaging. **a** ISAR two-dimensional image of a single metal corner reflector. **b** Power distribution along the azimuthal direction (dashed line in a) of the two-dimensional image. **c** and **d** Zoom-in views of two-dimensional image and the azimuthal slice, respectively.

**Performance comparison with previous photonic radar demonstrations:**
**Performance comparison**
Extended Data Table 1 shows a performance comparison of our work and previous photonic radar demonstrations, indicating significantly improved functionalities and performance metrics in terms of integration level, center frequency, bandwidth, distance resolution, velocity resolution, and imaging resolution.

**Extended Data Table. 1 Performance comparison of photonic microwave and mmWave radars**

| Ref. | Integration /Platform | Center frequency (GHz) | Bandwidth (GHz) | OSCR (dB) | Functionality | | |
|---|---|---|---|---|---|---|---|
| | | | | | Distance resolution (cm) | Velocity resolution (m/s) | Image resolution (cm×cm) |
| [5] | No | 36.0 | 8.0 | N/A | 2.00 | N/A | 3.30×1.90 |
| [8] | No | 39.8 | 0.2 | 25 | 2300.00 | 0.556 | N/A |
| [9] | No | 22.0 | 8.0 | 16 | 2.00 | N/A | 2.00×2.00 |
| [10] | No | 22.0 | 8.0 | 19.87 | 1.875 | N/A | N/A |
| [14] | No | 15.0 | 6.0 | N/A | 3.59 | N/A | N/A |
| [15] | No | 10.0 | 4.0 | N/A | N/A | N/A | N/A |
| [16] | No | 15.0 | 4.0 | N/A | 1.88 | N/A | 1.88×2.00 |
| [17] | No | 14.6 | 0.6 | N/A | 25.00 | N/A | 25.00×25.00 |
| [18] | No | 5.0 | 10.0 | N/A | 1.37 | N/A | N/A |
| [19] | No | 94.5 | 5.0 | 12.5 | 3.00 | N/A | N/A |
| [20] | No | 96.5 | 9.6 | N/A | 1.53 | N/A | N/A |
| [21] | No | 57.5 | 7.0 | 33 | 2.14 | N/A | 2.14×2.14 |
| [27] | Yes/SOI | 14.1 | 0.1 | N/A | 150.00 | N/A | N/A |
| [28] | Yes/SOI | 9.9 | 0.04 | N/A | 375.00 | 5.000 | N/A |
| [29] | Yes/SOI | 15.0 | 6.0 | 5 (25 with filter) | 2.70 | N/A | 2.70×2.70 |
| **This work** | **Yes/TFLN** | **45.0** | **10.0** | **30** | **1.50** | **0.067** | **1.50×1.06** |

N/A=Information not available or not applicable
OSCR= Optical sideband to carrier ratio

**Funding.**
Research Grants Council, University Grants Committee (CityU 11204820, CityU 11212721, CityU 11204022);

Croucher Foundation (9509005); National Natural Science Foundation of China (No. 62305008).

**Acknowledgments.**
We thank Dr. Wing-Han Wong and Dr. Keeson Shum at CityU for their help in device fabrication and measurement. We thank the technical support of Mr. Chun Fai Yeung, Mr. Shun Yee Lao, Mr. C W Lai and Mr. Li Ho at HKUST, Nanosystem Fabrication Facility (NFF) for the stepper lithography and PECVD process.

**Author contributions**
S.Z., N.Z. and C.W. conceived the idea. S.Z. proposed the system architecture and designed the devices. S.Z., Y. Z. and H.F. fabricated the wafer. S.Z. and Y.Z. carried out the experimental measurements. S.Z., J.F., Y.W. and K.Z. analyzed the data. S.Z. prepared the manuscript with contribution from all authors. E.P., N.Z. and C.W. supervised the project.

**Conflict of interest.** The authors declare no conflicts of interest.

**Data availability.** Data underlying the results presented in this paper are not publicly available at this time but may be obtained from the authors upon reasonable request.

**Code availability.** The code that supports the findings of this study is available from the corresponding authors upon reasonable request.